# Developing a Neural Network Machine Learning Interatomic Potential for Molecular Dynamics Simulations of La-Si-P Systems

Ling Tang[3], Weiyi Xia[1,4], Gayatri Viswanathan[1,2], Ernesto Soto[1,2], Kirill Kovnir[1,2], and Cai-Zhuang Wang[1,4,*]

[1]Ames National Laboratory, U.S. Department of Energy, Ames, IA 50011, United States.

[2]Department of Chemistry, Iowa State University, Ames, IA 50011, United States.

[3]School of Physics, Zhejiang University of Technology, Hangzhou, 310023, China

[4]Department of Physics and Astronomy, Iowa State University, Ames, IA 50011, United States

Corresponding author: * wangcz@ameslab.gov

## Abstract

While molecular dynamics (MD) is a very useful computational method for atomistic simulations, modeling the interatomic interactions for reliable MD simulations of real materials has been a long-standing challenge. In 2007, Behler and Perrinello first proposed and demonstrated an artificial neural network machine learning (ANN-ML) scheme, opening a new paradigm for developing accurate and efficient interatomic potentials for reliable MD simulation studies of the thermodynamics and kinetics of materials. In this paper, we show that an accurate and transferable ANN-ML interatomic potential can be developed for MD simulations of La-Si-P system. The crucial role of training data in the ML potential development is discussed. The developed ANN-ML potential accurately describes not only the energy versus volume curves for all the known elemental, binary, and ternary crystalline structures in La-Si-P system, but also the structures of La-Si-P liquids with various compositions. Using the developed ANN-ML potential, the melting temperatures of several crystalline phases in La-Si-P system are predicted by the coexistence of solid-liquid phases from MD simulations. While the ANN-ML model systematically underestimates the melting temperatures of these phases, the overall trend agrees with experiment. The developed ANN-ML potential is also applied to study the nucleation and growth of LaP as a function of different relative concentrations of Si and P in the La-Si-P liquid, and the obtained results are consistent with experimental observations.

## 1. Introduction

Molecular dynamics (MD) simulation, for which Prof. Parrinello has introduced many groundbreaking methodologies over the last 45 years, is a powerful computational tool in studying the thermodynamics and kinetics of real materials in atomistic detail and has been widely used in today's materials research [1,2]. However, a major bottleneck for the application of MD methods in materials simulations is the lack of accurate and efficient interatomic potentials. While *ab initio* MD can accurately describe the interatomic interactions in most materials, it is computationally too expensive to be used for large-scale and long-time simulations. In turn, while larger-size and/or longer-time MD simulations can be handled with low computational costs using empirical interatomic potentials, accurate and transferable empirical potentials are difficult to obtain, especially for systems containing 3 or more chemical elements.

Recently, machine learning (ML) has emerged as a promising strategy to tackle the challenges in constructing interatomic potentials that can balance the accuracy and efficiency requirements in MD simulations [3-24]. Among various ML approaches for interatomic potentials development, the artificial neural network (ANN) ML scheme first developed by Behler and Parrinello [3] and later improved by Zhang et al. [14], has been shown to be a very robust approach for generating accurate and efficient interatomic potentials. While there is considerable current interest in generating an "universal" neural network interatomic potential for systems with any combination of chemical elements in the periodic table, herein we focus on an in-deep exploration of developing accurate and transferable ANN-ML interatomic potentials for efficient and reliable MD simulation studies of phase stability and phase transformation in complex ternary systems, using the La-Si-P phase space as a case study.

Ternary La-Si-P compounds are promising materials for several technological applications. Silicon phosphides of transition and/or rare-earth metals can often crystallize in noncentrosymmetric (NCS) crystal structures [25-27]. In such materials, the absence of inversion symmetry and significant hybridization of *d*, *f*, and *p*-orbitals may promote a plethora of emergent properties such as unconventional superconductivity, topologically non-trivial quantum properties over large energy windows, and quasiparticle behavior [28-31]. All five reported ternary compounds in the La-Si-P phase space are semiconductors [32-34]. The noncentrosymmetric phases $LaSi_2P_6$ and $Pna2_1$ $LaSiP_3$ show potential as infrared (IR) nonlinear optical (NLO) materials [35], while polymorphs of $LaSiP_3$ exhibit low thermal conductivities, an important trait for thermoelectric materials [33]. These reports indicate La-Si-P compounds hold promise for energy-related applications and have motivated recent computational and ML searches for new La-Si-P ternary phases [36, 37]. Several stable and low-energy metastable La-Si-P ternary compounds have been predicted. However, the experimental synthesis of the predicted new compounds remains a challenge due to a lack of knowledge about the optimal synthetic conditions for the new compounds. Information from MD simulations about temperature effects on the relative thermodynamic stabilities among the predicted and competing structures, as well as the kinetics of phase formation and growth, would be highly desirable for the optimization synthetic parameters. Unfortunately, no interatomic potential is available for the reliable MD simulation of this ternary system due to the complexity of the interatomic interactions among La, Si, and P atoms.

For reliable MD description and prediction of phase stability and transformation as a function of temperature or pressure, the interatomic potential used in the simulations should be able to accurately describe the interactions among the atoms at a variety of bonding environments; this requires consideration not only near the equilibrium of various phases, but also for configurations far from the energy local minima and those near the transition states between different phases. Although there are some "universal" ANN-ML model trends cover many chemical elements [38-43], the accuracy of such potentials to predict the phase stability and transformation of materials is still far from practical use. In fact, developing ML interatomic potentials for reliable MD simulation studies of phase stability and phase transformation for ternary systems remains a challenging task. To develop an accurate and transferable ANN-ML interatomic potential for such MD simulation studies, it is critical to generate sufficient and effective training data covering not only the states near equilibrium, but also transient states across the transitions. In Ref. [44], Parrinello et al. showed that effective training data can be generated with the assistance of metadynamics simulations [45] (which was also pioneer-developed by Prof. Parrinello). In this paper, we will show that an accurate and transferable ANN-ML interatomic potential for MD simulations of La-Si-P ternary systems can also be achieved using an iterative approach by adaptively adding new training data (particularly those near the transition states) and adjusting the weight of existing data sets. After these two fine-tuning processes, the ANN-ML model well-reproduces the properties of the La-Si-P system, including crystals and liquids, compared with *ab initio* results. We also perform MD simulations to show the performance of the developed ANN-ML model in predicting the melting temperature and crystallization of the compounds.

The paper is organized as follows. In Section 2, we describe our iterative approach for developing an ANN-ML interatomic potential for the La-Si-P ternary system. The performance of the trained ANN-ML interatomic potential in MD simulations is presented in Section 3. Finally, summary discussions are given in Section 4.

## 2. An iterative approach for ANN-ML interatomic potential training

The ANN-ML interatomic potential for the La-Si-P system is trained using the deep learning software package DeePMD-kit [12-16]. A schematic illustration of the ANN for modeling interatomic potentials is shown in **Fig. 1**. To model the interatomic potential by ANN, the information fed to the input layer is a set of descriptors $D_i$, which are generated from the atomic coordinates of the input structures and a filter neural network based on the radial information of neighbors around each atom $i$. Thus, $D_i$ describes the atomistic environment around every atom $i$ of the structures in the training data set. The information extracted from the output layer is the energy $E_i$ on each atom. Then the total potential energy $E$ of each structure can be written as the sum of atomic energy $E_i$, i.e., $E = \Sigma_i E_i$. The mapping from the local environment of atom $i$ (i.e., $D_i$) to the energy of each atom $E_i$ is done by the hidden layers in the ANN. The forces on each atom can be readily obtained from the derivatives of the total energy. In all liquids and distorted crystals in training data set, the local structure configuration information within a cutoff radial of $R_{cutoff}$ = 7.0 Å is sampled to train the ANN-ML model with four hidden layers and 120 nodes per layer. The tanh function is used as an activation function in the neural network model.

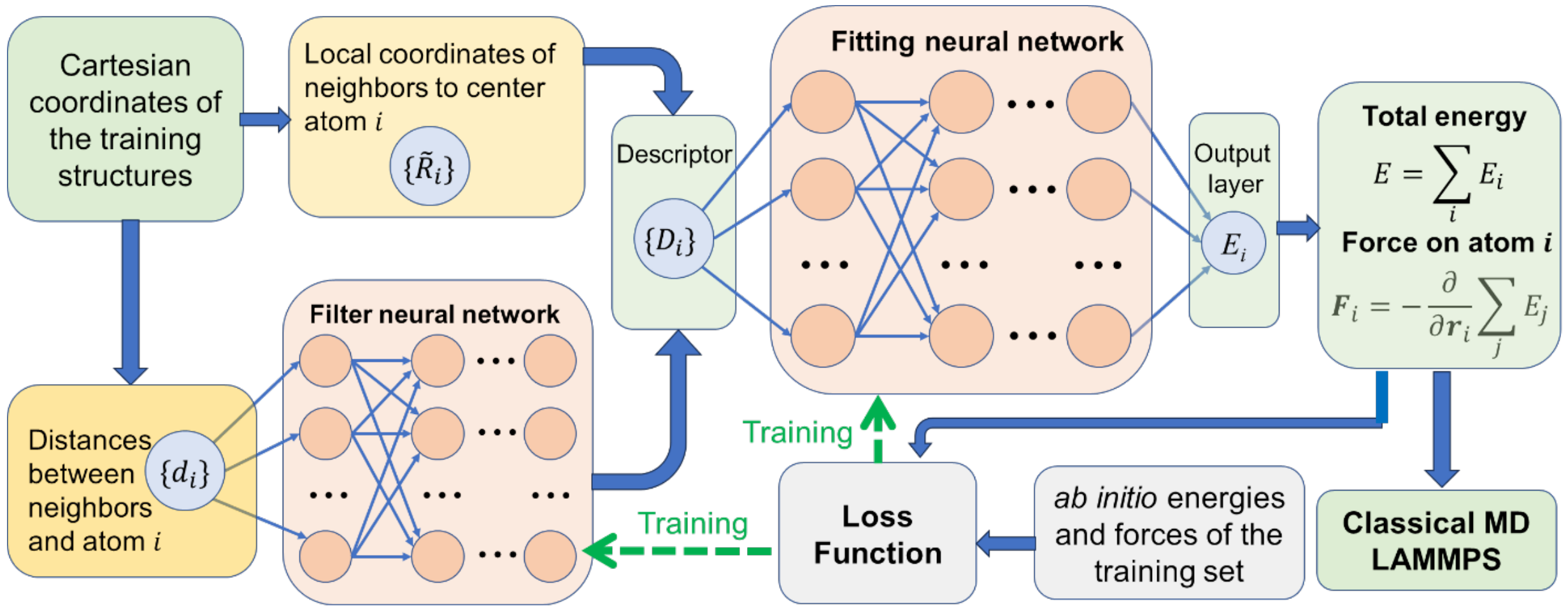

**Fig. 1.** Schematic illustration of the ANN for modeling interatomic potentials. The descriptors as an input layer of fitting the neural network are constructed by local atomic coordinates of neighbors to atom $i$ and the output layer of an embedded filter neural network, which provides weight factors of each neighbor according to the distances between the neighbors and atom $i$.

While the ANN deep learning software for interatomic potential development is readily available [13], the key challenge is how to develop proper training data sets to achieve accurate and transferable ANN-ML potentials. We generate initial training data that covers a wide range of compositions and bonding environments including distorted crystalline structures and liquid snapshots as shown in **Fig. 2 (a)**, where the liquid snapshots are generated by *ab initio* MD (AIMD) simulations. The red dots in **Fig. 2 (a)** indicate the compositions used for the distorted crystalline structures. The parent crystalline structures are the stable elemental, binary and the currently known ternary phases of La-Si-P system taken from Materials Project database [46], as well as several new ternary compounds predicted by recent ML studies [36, 37] including $La_2SiP$, $La_5SiP_3$, $La_2SiP_3$ (also denoted by the short notation 211, 513, 213, respectively). The recent experimental synthesized $La_2SiP_4$ (214) is also included in the parent. For each of these parent crystalline structures, 21 to 41 non-distorted structures are generated by uniformly expanding or contracting the volume of the initial structure, then 2100 to 4100 distorted structures are generated by randomly distorting atomic positions from their lattice position with a displacement amplitude of 0.2 Å. In this way, a total of 71,800 "distorted" crystal structures are generated for the ANN-ML training. These structures cover a wide range of compositions as can be seen from **Fig. 2 (a).** In addition to the distorted crystalline structures, liquid structures with various compositions including $La_xSi_yP_{100-x-y}$, $La_xSi_{100-x}$, $La_xP_{100-x}$, $Si_xP_{100-x}$ ($x$ or $y$ = 20, 40, 60, 80), $La_{10}Si_{80}P_{10}$, $La_5Si_{90}P_5$, $La_{10}Si_{10}P_{80}$, and $La_5Si_5P_{90}$, as well as pure La, pure Si, and pure P liquids are also generated by the AIMD simulations, as indicated by the blue dots in **Fig. 2 (a)**. For each of these liquids, 2,000 to 60,000 snapshot structures within a simulation time of 6 to 180 ps are extracted. A total of 130,000 liquid snapshot structures are therefore also included in the ANN-ML potential training. We then perform first-principles density-functional theory (DFT) calculations to obtain the total potential energy and forces on each atom for all these structures and use them in the training.

The *ab initio* calculations and AIMD simulations for the training data generation are performed using the VASP package [47, 48]. The core-valence electron interaction is described by the projector-augmented-wave (PAW) method. The exchange and correlation energy functional is non-spin-polarized generalized gradient approximation (GGA) in the Perdew–Burke–Ernzerhof (PBE) form [49]. The energy cutoff for the plane wave basis is 400 eV. In AIMD simulations of liquid phases, a canonical (NVT) ensemble with Nose-Hoover thermostat [50, 51] is used and the Brillouin zone is sampled only by the gamma point. The simulation box contains 256 atoms for each AIMD of liquid. The time step used in the AIMD simulation is 3 fs. DFT calculations to obtain the total energy and forces on each atom for each distorted crystal or liquid snapshot described above are performed with a k-point grid of $2\pi \times 0.02$ Å$^{-1}$.

**Fig. 2 (b)** shows the schematic illustration of our iterative scheme for the ANN-ML interatomic potential training. Using the initial training data described above, a pretrained ANN-ML model is developed. Since we initially did not know which phase (structure) is difficult for the ANN model to learn, all structures in the training data set have equal usage probability in the model learning process for the pretrained model. After obtaining the pretrained ANN-ML model, we randomly select some La-Si-P phases to test the fitting errors in the energies and forces. If the fitting error of a phase is larger than the others, the usage probability of this phase in model learning is increased in the next model training iteration. In general, after several iterations, an accurately trained ANN-ML model can be obtained. Then, we perform calculations and MD simulations using the obtained ANN-ML model to test the prediction of energy verses volume (E-V) curves for the crystalline structures and the pair-correlation functions for all liquid phases in the training data set. If the prediction of the E-V curve of a crystalline or g(r) of a liquid phase by the ANN-ML model has larger error than expected compared to DFT results, then more DFT data related to such crystalline or liquid phases are added into the training data set. After several fine-tune iterations of the ANN-ML model, we obtain the final accurate and transferable ANN-ML model for the La-Si-P system.

An example of improving the prediction of the E-V curve for crystalline phosphorus (P) by adjusting the usage probabilities of data set is shown in **Fig. 2 (c)**. During the iterative training process, we found the E-V curve of elemental P (which is # mp-1198724 in Materials Project database and all the following P refer to this polymorph of phosphorus) in the 4$^{th}$ iteration of training has larger error when the volume is expanded (top panel of **Fig. 2 (c)**). We then increase the probability of using P crystal data and slightly decrease the probability of the other systems (to maintain total probability normalization) in the next iterations of model training. After several training iterations with a continuous increasing in the probability of P data usage, the prediction accuracy of E-V curve for the P phase is significantly improved as shown in the bottom panel of **Fig. 2 (c)**. Another example is the improvement of melting point prediction for $La_2SiP_4$ by adding relevant crystal melting data into the training data, as shown in **Fig. 2 (d)**. From the initial model to the 3$^{rd}$ iteration model, the training data are the same as that described in the initial data generation described above, and only the data usage probability varies in the iterative training process. As one can see from the plot, the melting temperature of the 214 phase is affected very little between the initial to 3$^{rd}$ iteration training. In the 4$^{th}$ iteration and afterward, using the pretrained ANN-ML model to add to the training data set with additional structures collected during the crystal to liquid transition from the MD simulations, results in the melting temperature increasing substantially, as evident in **Fig. 2 (d)**. The solid/liquid coexistence MD simulation for the final model is shown in the following section and indicates that the final model, which includes solid-liquid transition configurations in the training data, predicts melting temperature in much better agreement with the experimental value determined by differential scanning calorimetry.

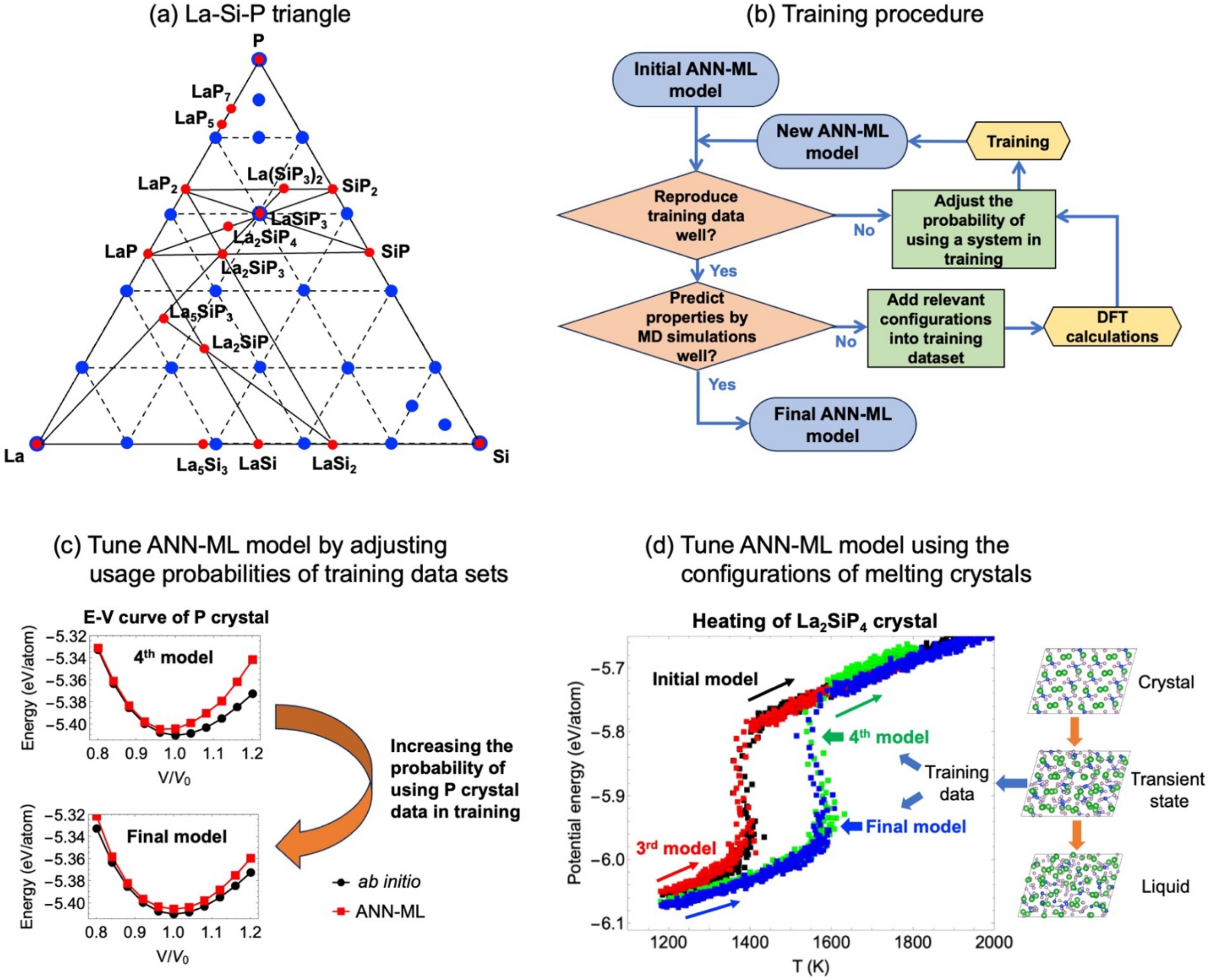

**Fig. 2. (a)** Ternary phase diagram highlighting the La-Si-P phases used in the ANN ML interatomic potentials training, where the compositions of crystalline and liquid structures in training data set are indicated by red and blue dots, respectively. **(b)** A schematic illustration of the iterative training process dictating the workflow with several fine-tune training iterations. **(c)** An example of improvement of E-V curve prediction for crystalline P by adjusting the usage probabilities of the data set. **(d)** An example of improvement of melting temperature prediction for $La_2SiP_4$ by adding relevant training data from MD simulation of liquid to solid transitions.

The training accuracy of the final ANN-ML interatomic potential is shown in **Fig. 3**. The root mean square (RMS) error of the energy is very small (~12 meV/atom) while the error in the force is about 0.23 eV/Å. Such accuracy is acceptable for MD simulations.

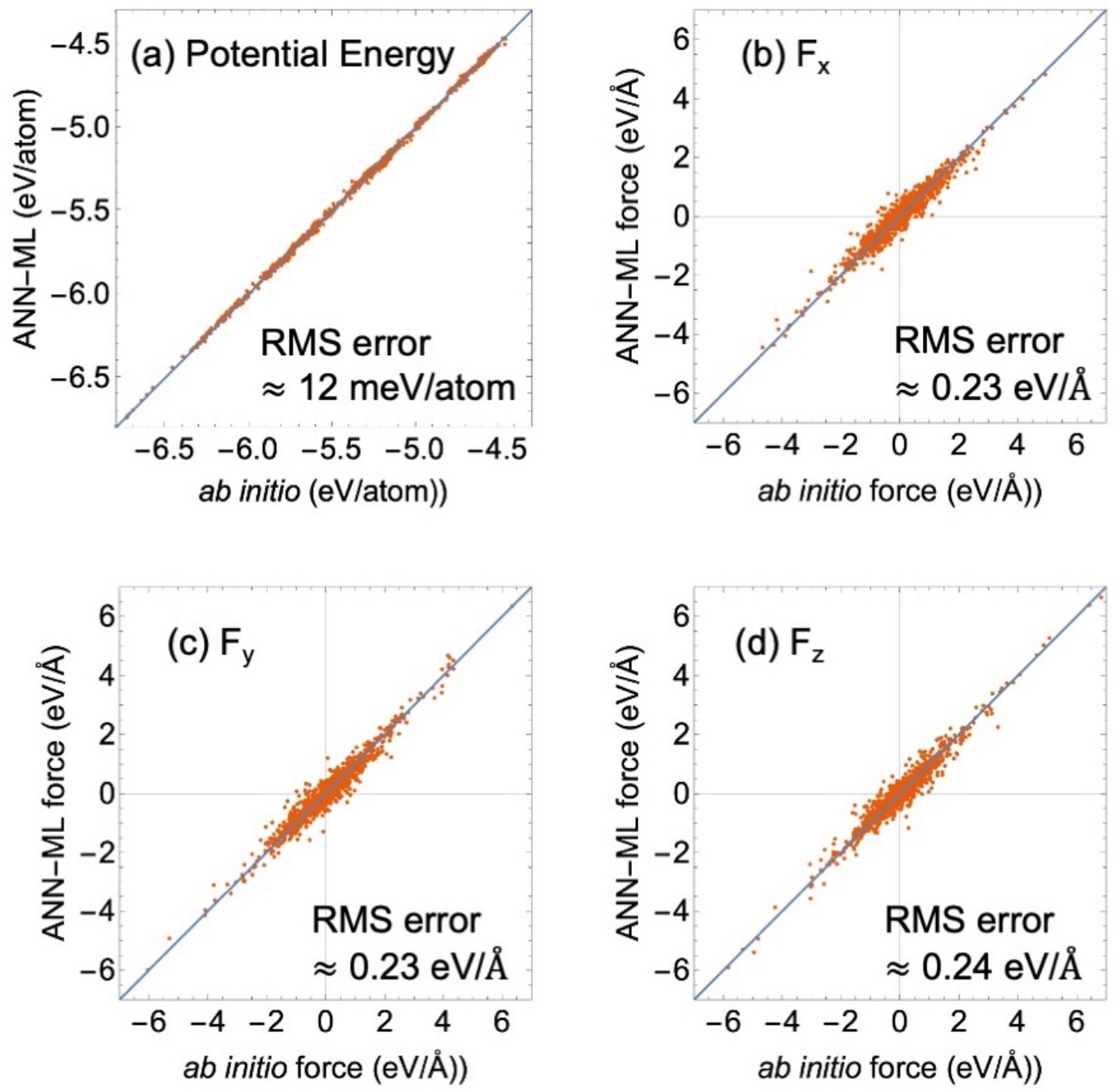

**Fig. 3.** Potential energy **(a)** and force **(b)-(d)** predictions by the final ANN-ML model are compared with *ab initio* calculation results. The comparison of the x, y, and z components of forces are shown in **(b)-(d)** respectively. Here, we randomly selected 20 snapshots for each system (total 66 systems) in training data set and compare their ANN-ML predicted energy and forces with *ab initio* results. The RMS error calculation is based on these randomly selected 1320 data in the training data set.

## 3. Performance of the developed ANN-ML potential

The accuracy and transferability of the developed ANN-ML interatomic potential for predicting structures and properties in the La-Si-P system are further evaluated by the comparing the ML potential calculation/simulation results with those from *ab initio* calculations and AIMD simulations on the energy versus volume curves for various crystalline phases (**Fig. 4**) and the structures of liquid phases at different compositions (**Fig. 5**). Here, the structure relaxations and MD simulations using ANN-ML potential are performed using the LAMMPS package [52], while the *ab initio* calculations are performed using the VASP package.

We used the crystal structures of all known stable ($E_{hull}$ = 0) elemental, binary, and ternary phases in the La-Si-P system from the Materials Project database [46] and some ternary phases from the recent ML predictions [36, 37] to assess the transferability of the developed ANN-ML interatomic potential. As can be seen from **Fig. 4**, our ANN-ML potential produces E-V curves in good agreement with the results from *ab initio* calculations. In **Fig. 5**, we show the total pair-correlation functions for all liquid phases (indicated by the blue dots in **Fig. 2 (a)**) obtained by AIMD and ANN-ML MD simulations at 2500 K. The ANN-ML potential also well reproduces the atomic structures of La-Si-P liquids as compared with the *ab initio* results.

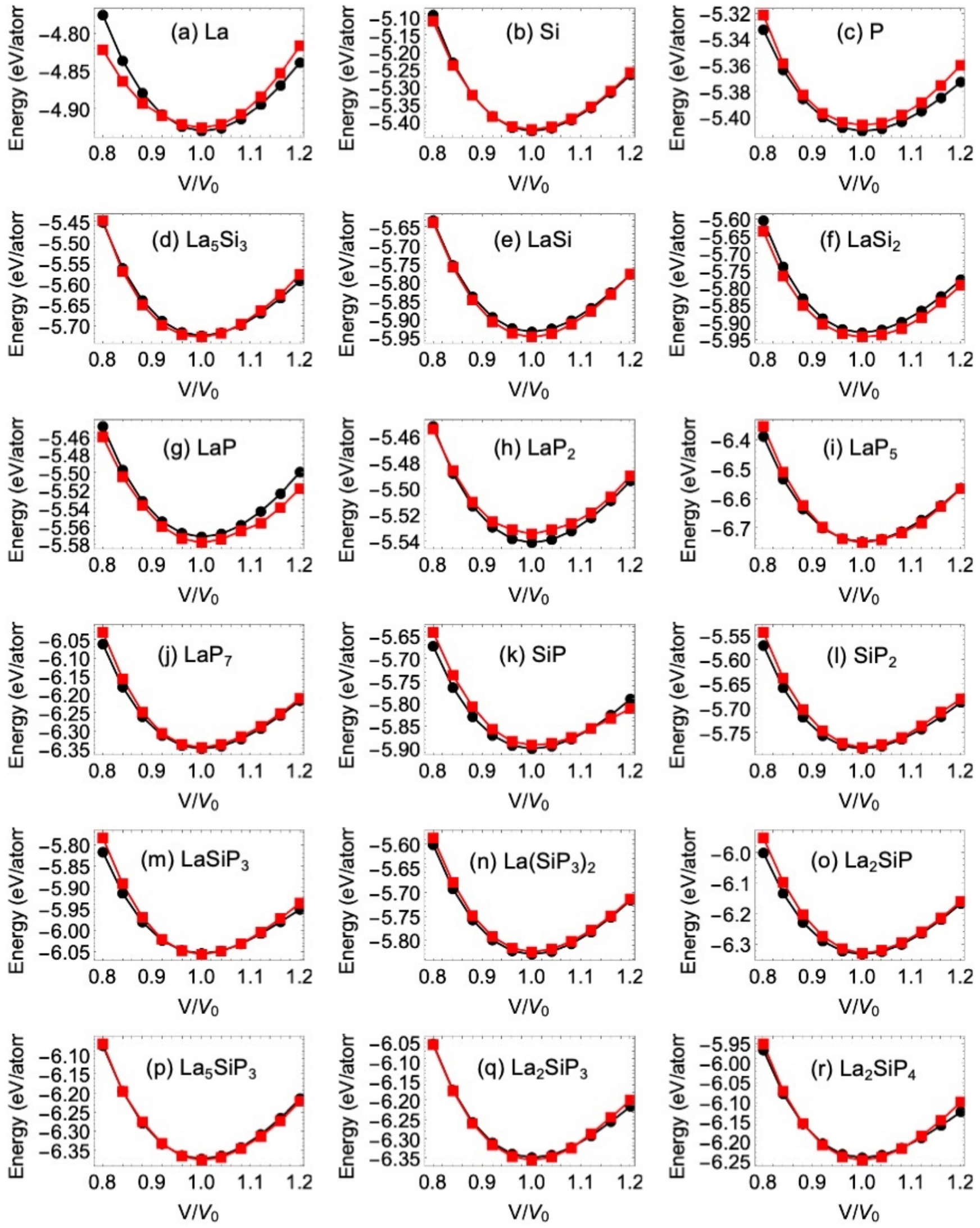

**Fig. 4.** Comparisons of E-V curves of crystalline La-Si-P phases calculated by the *ab initio* method (black) and ANN-ML potential (red).

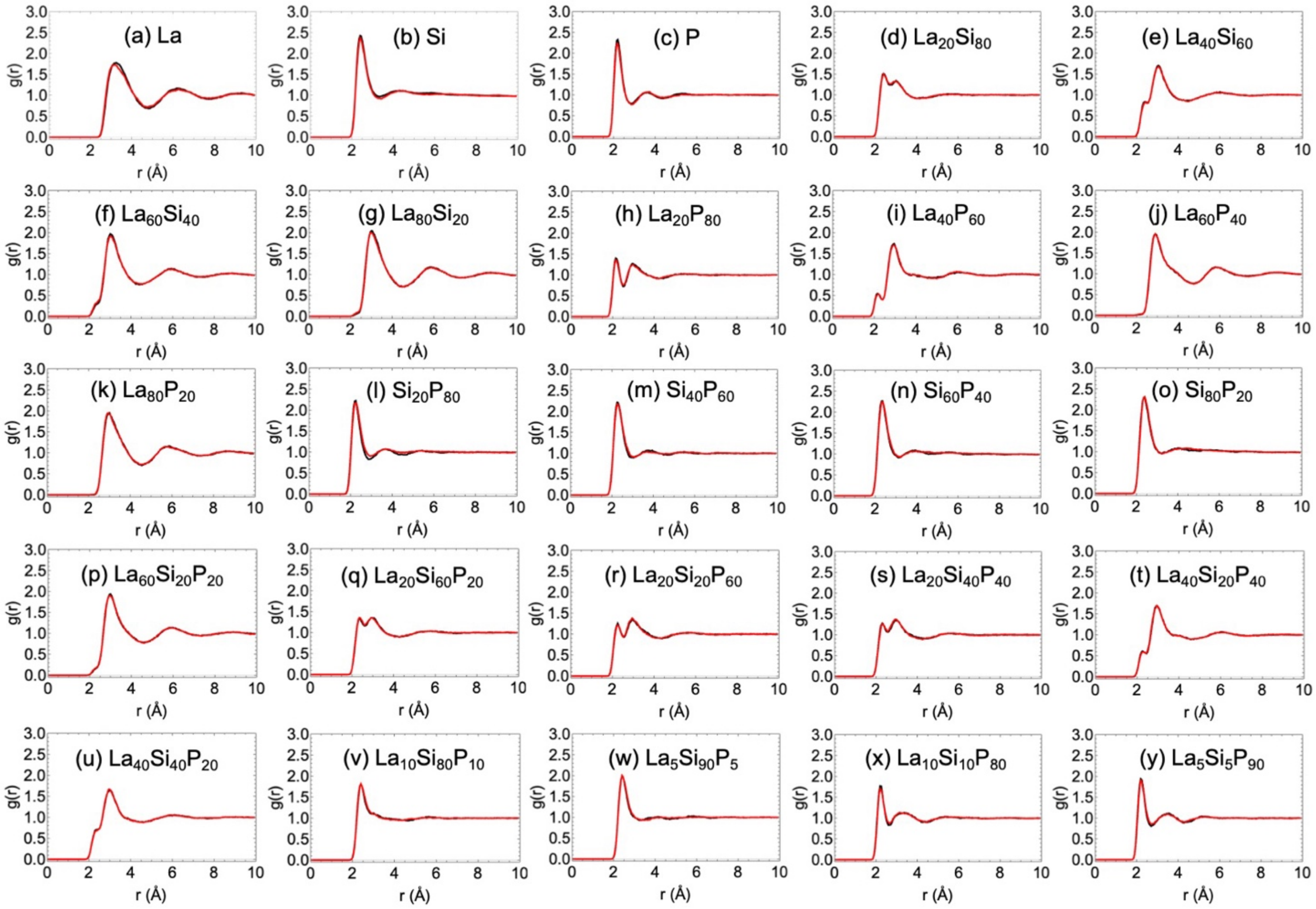

**Fig. 5.** Comparison of the calculated total pair-correlation functions from *ab initio* MD (black) and ANN-ML MD simulations (red) for La-Si-P liquid phases. All the MD simulations are in the NVT ensemble and the temperature used for the simulations is 2500K.

Using the developed ANN-ML interatomic potential for La-Si-P systems, we also performed MD simulations to predict the melting temperatures of the ternary $LaSiP_3$ and $La_2SiP_4$ phases by the solid/liquid coexistence method [53]. The initial MD simulation box is a crystalline supercell. An isothermal-isobaric (NPT) ensemble and a Nose-Hoover thermostat [50, 51] are used to prepare a solid/liquid interface. First, the atoms in the whole simulation box are allowed to move for 10 ps at a temperature slightly below the estimated melting point. Then, while the atoms in the left half of box remain frozen at the snapshot at the end of 10 ps, the atoms in the right half of box are continuously heated up to 2600 K followed by cooling down to the initial simulation temperature when a complete melting is observed. Finally, a 10 ps NPT simulation is performed for all atoms to prepare the initial configurations of solid/liquid interface.

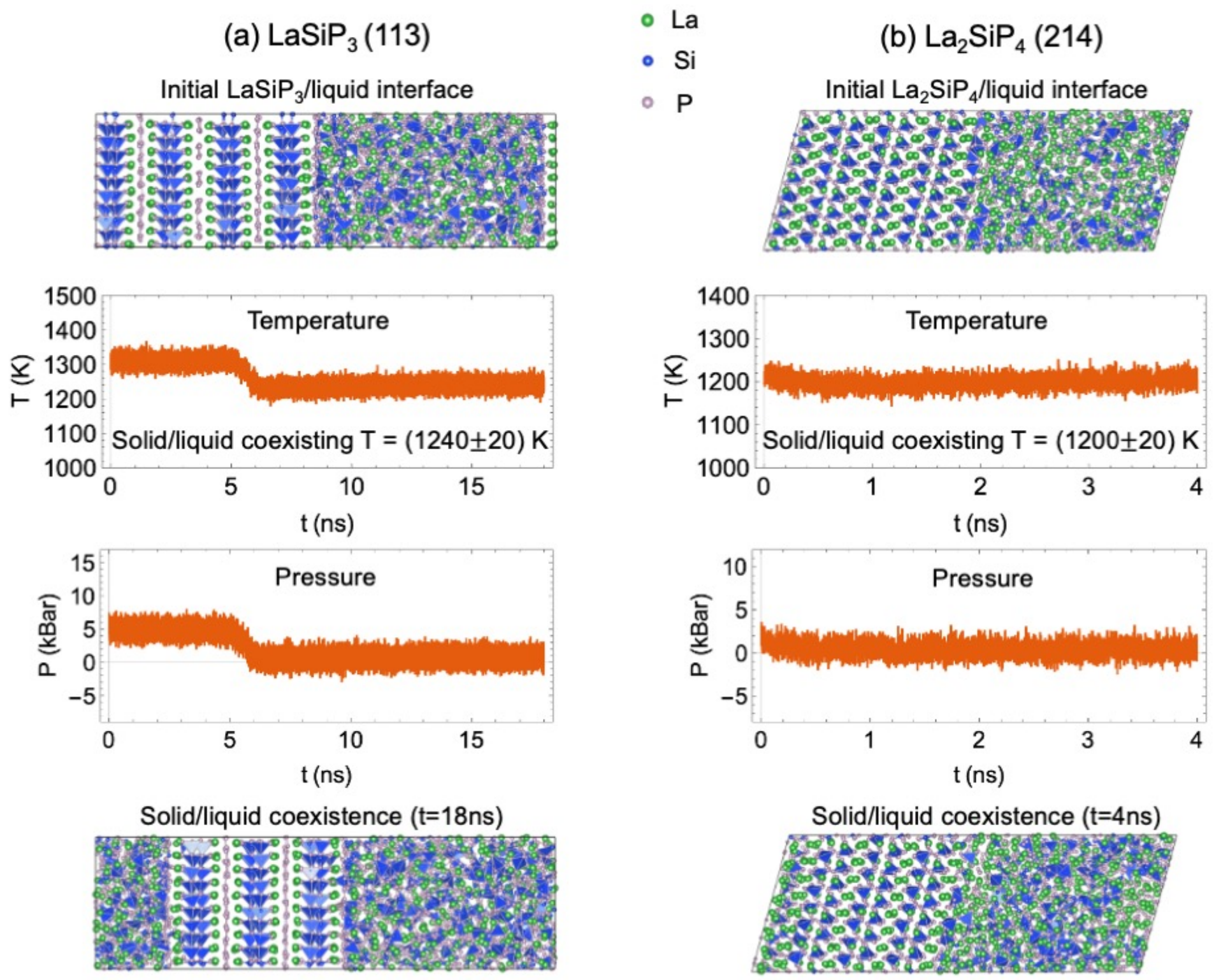

**Fig. 6.** Solid/liquid coexistence simulation of **(a)** *Aea*2 $LaSiP_3$ and **(b)** $La_2SiP_4$ ternary phases. The top panels show the initial solid/liquid interface structure used in the simulation and the bottom panels show the coexistence of the solid and liquid phases. The two middle panels show the time evolution of the temperature and pressure, respectively, when the total energy in the NVE simulation is set to -5.544 eV/atom ($LaSiP_3$) and -5.796 eV/atom ($La_2SiP_4$).

Starting from the solid/liquid interface structure described above, the solid/liquid coexistence MD simulation is performed using a NVE ensemble (i.e., the number of particles, the volume of the simulation box, and the total energy of the system are kept constant). Coexistence is searched by adjusting the total energy of the systems through scaling the velocities of the atoms. Once the total energy is too high or too low, the whole system will completely melt or crystallize. If an appropriate total energy is chosen, then the coexistence of crystalline and liquid phases can be reached, and the melting temperatures can then be determined by averaging the kinetic energy of the system over a sufficient time in the MD simulation. The coexistence calculations between the solid and liquid phases near zero pressure for the ternary *Aea*2 $LaSiP_3$ (which is not included in Materials Projection database but whose structure and melting temperature have been obtained by experiment in Ref. [33]) and $La_2SiP_4$ phases are shown in **Fig. 6 (a)** and **(b)** respectively. The melting temperatures are evaluated to be 1240±20 K and 1200±20 K, respectively, for the *Aea*2 $LaSiP_3$ and $La_2SiP_4$ phases around zero pressure.

From literature, we find experimental value for the melting temperature of *Aea*2 $LaSiP_3$ phase is 1308 K [33]. The predicted value from our MD simulation using the developed ANN-ML interatomic potential is about 5.2% lower than the experimental value. For the $La_2SiP_4$ phase, there is no experimental data on its melting temperature in the literature. To determine the experimental melting temperature for this ternary phase, we conducted a differential scanning calorimetry (DSC) experiment for the $La_2SiP_4$ phase by sealing a small amount of powder in an evacuated silica ampoule and heating/cooling the sample to 1373K. The result of this experiment is shown in **Fig. 7**. Upon heating, $La_2SiP_4$ melts at ~1330 K which is about 9.8% higher than the predicted melting point, $T_m$, of 1200 K using the crystal/liquid coexistence MD simulation method. It is possible that the crystal-liquid coexistence temperature of $La_2SiP_4$ could be affected by P vapor pressure and the conditions of the DSC experiment. The melting temperature of GeAs was different when measured under a saturated arsenic environment (equilibrium melting temperature) as opposed to a sealed, evacuated ampoule [54]. In a simple binary system, this resulted in ~20 K difference in $T_m$ due to the presence of pnictogen vapor pressure, indicating that the experimental $T_m$ of pnictides is dependent on measurement conditions. Comparison of experimental data and our simulation results indicate that the ANN-ML potential systematically underestimates the melting temperature of La-Si-P ternary compounds by about 5-10%. We note that the melting temperatures are not explicitly included in training the ANN-ML interatomic potential, thus the accuracy of the ANN-ML interatomic potential in predicting the melting temperatures of these complex ternary compounds is remarkable.

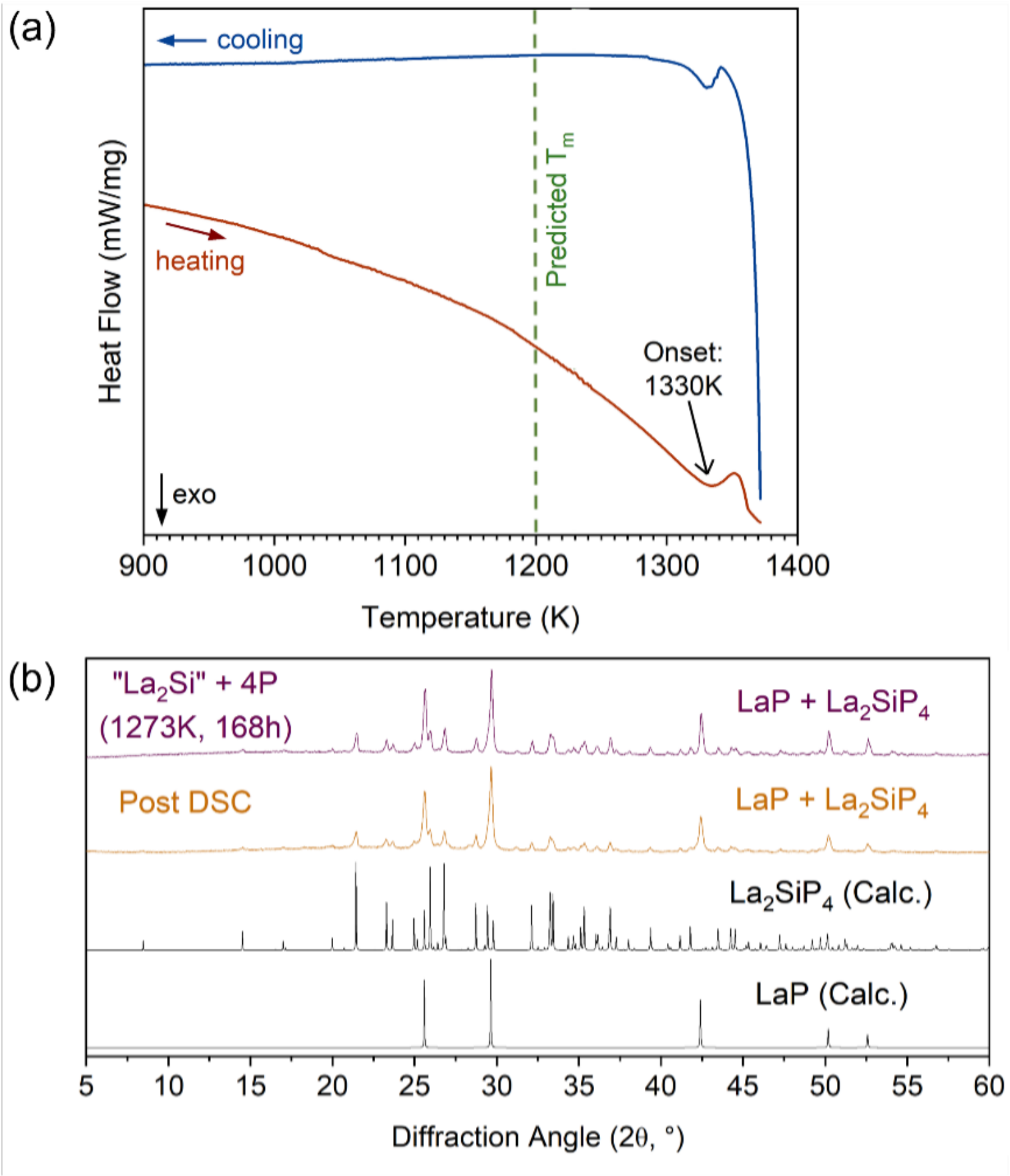

**Fig. 7.** Thermal stability of $La_2SiP_4$. **(a)** Differential scanning calorimetry (DSC) curves for heating and cooling a sample of (LaP + $La_2SiP_4$) up to 1373 K. An endothermic peak is evident upon heating at ~1330 K, likely corresponding to the melting of $La_2SiP_4$, followed by an exothermic peak on cooling which indicates recrystallization. The predicted crystal-liquid coexistence temperature, $T_m$, is indicated by the vertical dashed line. **(b)** Powder X-ray diffraction (PXRD) patterns of the sample before (purple) and after (orange) DSC treatment together with calculated patterns for each phase (black curves). Before DSC, LaP is the majority phase with minor $La_2SiP_4$. No significant change is observed after DSC.

To further test the transferability of the developed La-Si-P ANN-ML interatomic potential, we performed solid/liquid coexistence MD simulations to predict the melting temperatures for binary (LaSi and LaP), and elemental (La and Si) crystalline phases. The simulation procedure is the same as that used for the ternary phase simulations described above, except that heating to a higher temperature (3400 K) is needed to generate the liquid for LaP binary phase in the right half of the simulation box. **Fig. 8 (a)** shows the predicted melting temperature $T_m$ for LaSi phase is ~1500 K which is about 20% lower than the experimental value 1893 K [55]. For LaP phase, our predicted melting temperature is about 2690 K as shown in **Fig. 8 (b)**. This predicted value is very high and there is no literature report for the thermal stability of LaP. Our attempt to determine the melting temperature of LaP using DSC revealed no endothermic or exothermic peaks on heating/cooling up to the maximum temperature of 1373K, supporting our predictions. For the elemental phases, the predicted $T_m$ for La is about 960 K as shown in **Fig. 8 (c)** which is 19% lower than the experimental value 1191 K [55], and the predicted $T_m$ for Si is ~1520 K as shown in Fig. 8 (d) which is 10% lower than the experimental value 1687 K [55]. It is interesting to note that our predicted $T_m$ for Si is close to the predicted $T_m$ by GGA-DFT [56], which also underestimated the melting point of Si by about 10%. Since our training data is generated by GGA-DFT calculations, the systematic underestimation of melting temperature by our developed ANN-ML interatomic potential in comparison with experimental data would be partially attributed to the use of DFT data in the ML training.

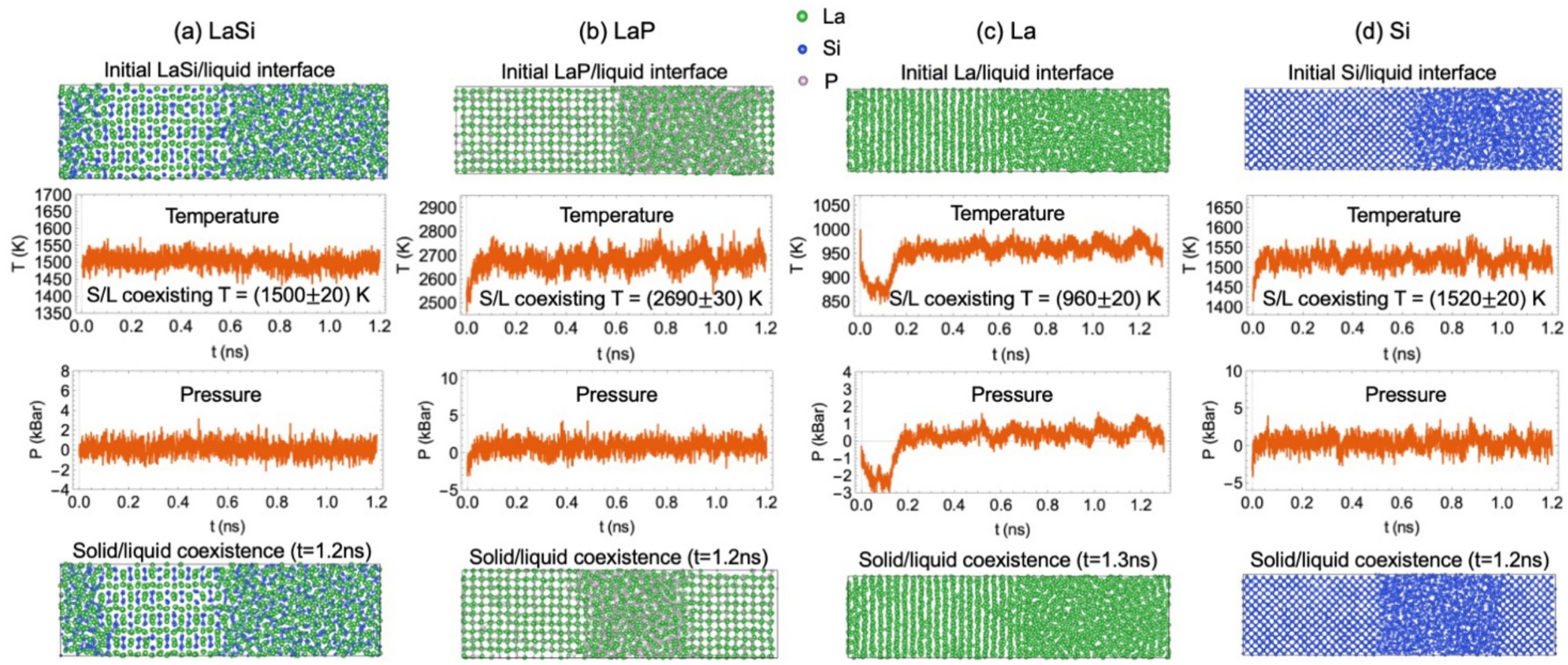

**Fig. 8.** Solid/liquid coexistence simulations of **(a)** binary LaSi, **(b)** binary LaP, **(c)** elemental La, and **(d)** elemental Si phases. The top panel shows the initial solid/liquid interface structure used in the simulation and the bottom panel shows the coexistence of the solid and liquid phases. The two middle panels show the time evolution of the temperature and pressure, respectively, when the total energies in the NVE simulation are properly chosen to observe the solid/liquid coexistence.

The developed ANN-ML potential is also very useful for performing MD simulations to acquire information about phase competition, formation, and growth kinetics to understand the challenges in experimental synthesis of ternary La-Si-P phases [57]. Here we show the MD simulations using the developed ANN-ML potential for the nucleation and growth of LaP as function of different relative concentrations of Si and P in the liquid phase. The simulations are performed using an NPT ensemble and an 8×8×8 LaP supercell (4096 atoms). First, the simulations are performed at high temperature (T=3600 K for LaP liquid and T=2600 K for La-Si-P liquids) to ensure the lattice is melted to the liquid state. After isothermal simulation at high temperature for 200 ps, all the liquids are quenched to 1400 K and then the isothermal simulation is performed. As shown in **Fig. 9**, the potential energy has a sharp drop indicating a structure transition taking place. The atomic structures in the simulation box before and after the transition are shown in the insets of **Fig. 9**. The structures of LaP or Si-substituted LaP phases can be clearly seen in both the inset snapshot structures and the pair correlation functions shown in the bottom panel of each plot. As the Si concentration increases, more deviation from the perfect LaP structure is observed. These simulation results agree well with observations from experimental syntheses [57].

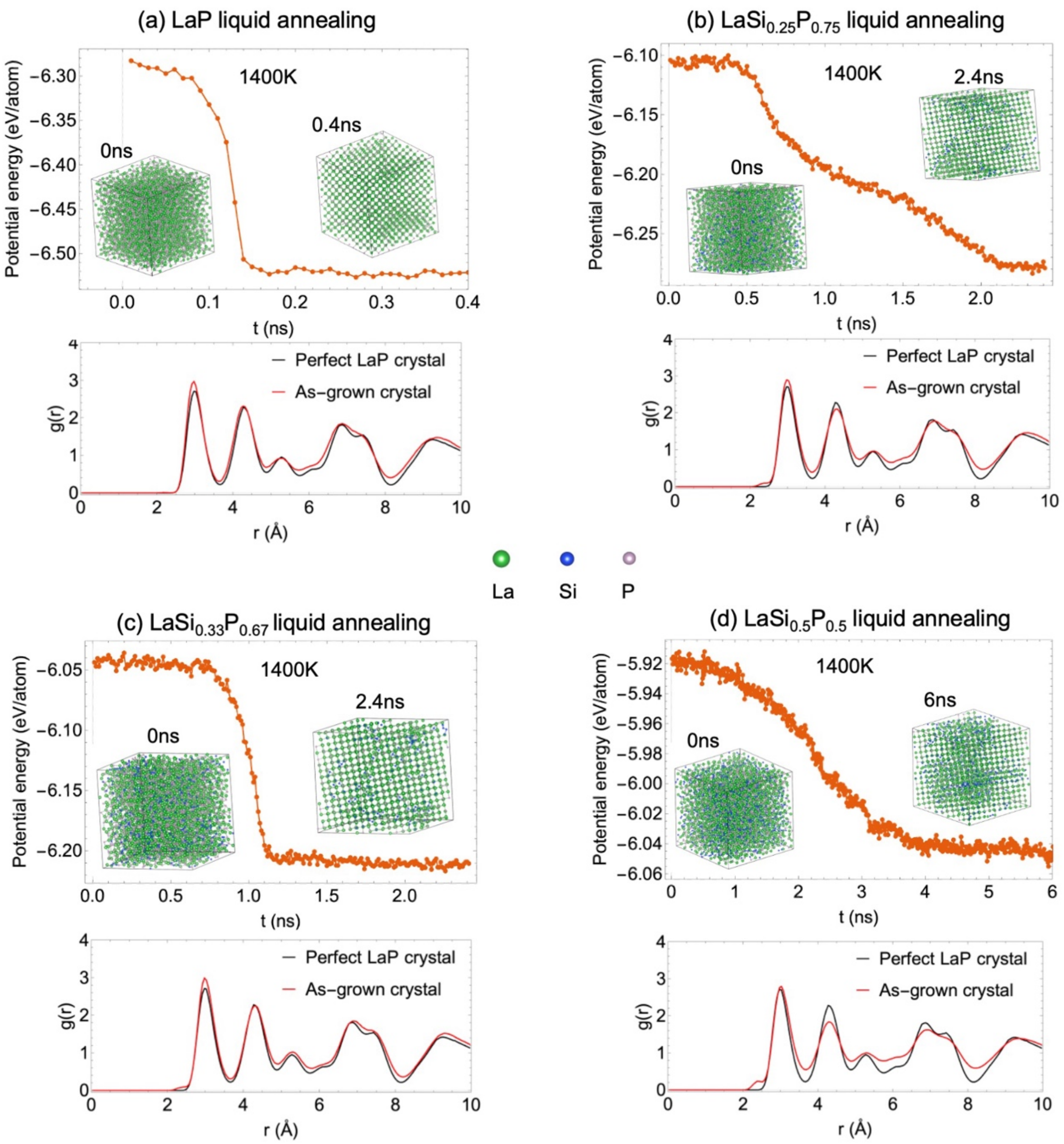

**Fig. 9.** Crystallization of LaP phases from undercooled liquids with varying concentrations of Si and P in the La-Si-P liquids. **(a)-(d)** Crystallization from $LaSi_xP_{1-x}$ liquid with $x$=0, 0.25, 0.33, and 0.5, respectively. The supercooled $LaSi_xP_{1-x}$ liquid at T=1400 K is prepared by quenching the high temperature liquid. Then, after isothermal annealing MD simulations, the LaP crystals are grown in the liquids as shown in the inset snapshots and the total pair-correlation functions at the bottom of each figure. The MD simulations show that the LaP crystals nucleate rapidly in the La-Si-P liquid, suggesting that the nucleation barrier is very low for the LaP phase.

## 4. Summary

In this paper, we explore the development of an ANN-ML interatomic potential for reliable MD simulation of the La-Si-P ternary system based on the ANN-ML scheme pioneer developed by Behler and Parrinello [3] and later improved by Zhang et al. [14]. We show that a comprehensive training data set covering a wide range of compositions and structures and an iterative training strategy are critical to obtain an accurate and transferable ANN-ML potential model. The atomistic structures in our training data set include not only the configurations near equilibrium, but also those far from equilibrium and the transient states across structural transitions. Our iterative training involves adaptively adjusting the usage probability of the training data set, and adding new structures and configurations to the training data set according to the performance of the ANN-ML potential obtained from previous training iterations.

Adjusting the usage probability of the training data set from one iteration to another can be done very efficiently. For each iteration, the parameters (4×120 nodes) in ANN-ML model are initialized by the last ANN-ML model, thus the training step is only about 1/3 of the pretrained process to ensure the convergence. By several training iterations of adjustment probabilities, the obtained ANN-ML model can well reproduce E-V curves for all studied crystalline phases in the La-Si-P system when compared with the DFT results. Meanwhile, the MD simulations by ANN-ML potential also well-reproduce the DFT calculated total pair-correlation functions for liquids.

Adding new structures/configurations to the training data set would require more experience. Bonati and Parrinello [44] use the configuration generated by metadynamics simulations to train a ANN-ML model to improve the performance of the model for predicting rare events such as nucleation. In the present study, we focus on crystal growth and melting during the materials synthesis. Therefore, we also add structures near the melting temperature to fine-tune the ANN-ML model. We use the pretrained ANN-ML model to generate snapshots close to the crystal melting. Then, we calculate the energy and forces of these snapshots using DFT. We found that adding these new training data can improve the performance of the ANN-ML model significantly, especially for melting temperature prediction.

We note that melting temperature prediction is a stern test for the developed interatomic potential since the melting temperature, $T_m$, is not explicitly included in the training data. The ANN-ML interatomic potential developed from our study performs well in predicting the melting temperatures of experimentally synthesized $LaSiP_3$ and $La_2SiP_4$ ternary compounds, with 5-10% underestimation. Moreover, the ANN-ML interatomic potential also correctly describes the crystallization of LaP and Si-substituted LaP phases from La-Si-P liquids, in good agreement with experiment observations. Since the main purpose of this potential development is to acquire thermodynamics and phase formation kinetics to efficiently guide the experimental synthesis of new ternary La-Si-P compounds, the performance of the ANN-ML potential for MD simulations is acceptable. The systematic error in predicting the melting temperatures is partially attributed to the use of DFT data in the ML training, which also tends to underestimate melting temperature relative to experimental data.

It is also interesting to note that the developed ANN-ML interatomic potential substantially (~20%) underestimates the melting temperature for some binary and elemental phases (i.e., LaSi and pure La) in comparison with the values obtained from experimental measurements. Although the use of DFT data in the ML training may contribute to some of the errors, more training data specifically for the melting transitions in these phases may be needed to improve the transferability of the potential. This can be a topic of future investigation.

**Acknowledgements**

Work at Ames National Laboratory was supported by the U.S. Department of Energy (DOE), Office of Science, Basic Energy Sciences, Materials Science and Engineering Division, including a grant of computer time at the National Energy Research Supercomputing Center (NERSC) in Berkeley. Ames National Laboratory is operated for the U.S. DOE by Iowa State University under contract # DE-AC02-07CH11358. We thank Dr. Yaroslav Mudryk (Ames National Laboratory) for access to the arc-melting setup. L. Tang acknowledges the support by the National Natural Science Foundation of China (Grant No. 11304279).

**Data availability**

The data that support the findings of this study are available from the corresponding author upon reasonable request.

**Conflict of interest**

The authors declare no competing interests.

**References**

[1] D. C. Rapaport, The Art of Molecular Dynamics Simulation, Cambridge University Press, Combridge (1995).

[2] Kun Zhou and Bo Liu, Molecular Dynamics Simulation: Fundamentals and Applications, Academic Press, New York (2022).

[3] J. Behler, M. Parrinello, Generalized Neural-Network Representation of High-Dimensional Potential-Energy Surfaces, *Phys. Rev. Lett.* **98**, 146401(2007).

[4] A. P. Bartók, M. C. Payne, R. Kondor, G. Csányi, Gaussian Approximation Potentials: The Accuracy of Quantum Mechanics, without the Electrons, *Phys. Rev. Lett.* **104**, 136403(2010).

[5] S. Chmiela, A. Tkatchenko, H. E. Sauceda, I. Poltavsky, K. T. Schütt, K.-R. Müller, Machine learning of accurate energy-conserving molecular force fields, *Sci. Adv.* **3**, e1603015(2017).

[6] A. V. Shapeev, Moment Tensor Potentials: A Class of Systematically Improvable Interatomic Potentials, *Multiscale Modeling & Simulation* **14**, 1153-1173(2016).

[7] I. S. Novikov, K. Gubaev, E. V. Podryabinkin, A. V. Shapeev, The MLIP package: moment tensor potentials with MPI and active learning, Machine Learning: *Science and Technology* **2**, 025002(2020).

[8] Z. Fan, W. Chen, V. Vierimaa, A. Harju, Efficient molecular dynamics simulations with many-body potentials on graphics processing units, *Comput. Phys. Commun.* **218**, 10-16(2017).

[9] Z. Fan, Z. Zeng, C. Zhang, Y. Wang, K. Song, H. Dong, Y. Chen, T. Ala-Nissila, Neuroevolution machine learning potentials: Combining high accuracy and low cost in atomistic simulations and application to heat transport, *Phys. Rev. B* **104**, 104309(2021).

[10] Y. Zhang, J. Xia, B. Jiang, Physically Motivated Recursively Embedded Atom Neural Networks: Incorporating Local Completeness and Nonlocality, *Phys. Rev. Lett.* **127**, 156002(2021).

[11] Y. Zhang, J. Xia, B. Jiang, REANN: A PyTorch-based end-to-end multi-functional deep neural network package for molecular, reactive, and periodic systems, *J. Chem. Phys.* **156**, 114801(2022).

[12] J. Han, L. Zhang, R. Car, and Weinan E, Deep potential: a general representation of a many-body potential energy surface, *Communications in computational physics* **23**, 629-639(2018).

[13] H. Wang, L. Zhang, J. Han, Weinan E, DeePMD-kit: A deep learning package for many-body potential energy representation and molecular dynamics, *Comput. Phys. Commun.* **228**, 178-184(2018).

[14] L. Zhang, J. Han, H. Wang, R. Car, Weinan E, Deep Potential Molecular Dynamics: A Scalable Model with the Accuracy of Quantum Mechanics, *Phys. Rev. Lett.* **120**, 143001(2018).

[15] T. Wen, L. Zhang, H. Wang, Weinan E, D. J. Srolovitz, Deep potentials for materials science, *Materials Futures* **1**, 022601(2022).

[16] L. Zhang, J. Han, H. Wang, W. Saidi, R. Car, End-to-end symmetry preserving inter-atomic potential energy model for finite and extended systems, Advances in Neural Information Processing Systems 2018, 31.

[17] L. Tang, K. M. Ho, C. Z. Wang, Molecular dynamics simulation of metallic Al-Ce liquids using a neural network machine learning interatomic potential, *J. Chem. Phys.* **155**, 194503(2021).

[18] L. Tang, Z. J. Yang, T. Q. Wen, K. M. Ho, M. J. Kramer, C. Z. Wang, Development of interatomic potential for Al–Tb alloys using a deep neural network learning method, *Phys. Chem. Chem. Phys* **22**, 18467-18479(2020).

[19] L. Tang, Z. J. Yang, T. Q. Wen, K. M. Ho, M. J. Kramer, C. Z. Wang, Short- and medium-range orders in Al90Tb10 glass and their relation to the structures of competing crystalline phases, *Acta Mater.* **204**, 116513(2021).

[20] J. Wang, H. Shen, R. Yang, K. Xie, C. Zhang, L. Chen, K. -M. Ho, C. -Z. Wang, S. Wang, A deep learning interatomic potential developed for atomistic simulation of carbon materials, *Carbon* **186**, 1-8(2022).

[21] J. Wu, Y. Zhang, L. Zhang, S. Liu, Deep learning of accurate force field of ferroelectric HfO2, *Phys. Rev. B* **103**, 024108(2021).

[22] C. Zhang, Y. Sun, H. -D. Wang, F. Zhang, T. -Q. Wen, K. -M. Ho, C. -Z. Wang, Crystallization of the P3Sn4 Phase upon Cooling P2Sn5 Liquid by Molecular Dynamics

Simulation Using a Machine Learning Interatomic Potential, *J. Phys. Chem. C* **125**, 3127-3133(2021).

[23] C. Zhang, L. Tang, Y. Sun, K. -M. Ho, R. M. Wentzcovitch, C. -Z. Wang, Deep machine learning potential for atomistic simulation of Fe-Si-O systems under Earth's outer core conditions, *Phys. Rev. Mater.* **6**, 063802(2022).

[24] T. Q. Wen, C. Z. Wang, M. J. Kramer, Y. Sun, B. L. Ye, H. D. Wang, X. Y. Liu, C. Zhang, F. Zhang, K. -M. Ho, and N. Wang, Development of a deep machine learning interatomic potential for metalloid-containing Pd-Si compounds, *Phys. Rev. B* **100**, 174101(2019).

[25] S. J. Lee, S. L. Carnahan, G. Akopov, P. A. Yox, L.-L. Wang, A. J. Rossini, K. Wu, K. Kovnir, Noncentrosymmetric Tetrel Pnictides RuSi4P4 and IrSi3P3: Nonlinear Optical Materials with Outstanding Laser Damage Threshold, *Adv. Funct. Mater.* **31**, 2010293(2021).

[26] S. J. Lee, G. Viswanathan, S. L. Carnahan, C. P. Harmer, G. Akopov, A. J. Rossini, G. J. Miller, K. Kovnir, Add a Pinch of Tetrel: The Transformation of a Centrosymmetric Metal into a Nonsymmorphic and Chiral Semiconductor, *Chem. Eur. J.* **28**, e202104319(2022).

[27] Ernesto Soto, Shannon J. Lee, Andrew P. Porter, Gayatri Viswanathan, Georgiy Akopov, Nethmi Hewage, Kui Wu, Victor Trinquet, Guillaume Brunin, Geoffroy Hautier, Gian-Marco Rignanese, Aaron J. Rossini, and Kirill Kovnir, FeSi4P4 and CoSi3P3: Hidden Gems of Ternary Tetrel Pnictides with Outstanding Nonlinear Optical Properties, *Chemistry of Materials* **36**, 8854-8863(2024).

[28] E. Bauer, M. Sigrist, Non-Centrosymmetric Superconductors: Introduction and Overview. Heidelberg, Ger.: Springer (2012).

[29] E. M. Carnicom, W.Xie, T. Klimczuk, J. Lin, K. Górnicka, Z. Sobczak, N. P. Ong, R. J. Cava, TaRh2B2 and NbRh2B2: Superconductors with a chiral noncentrosymmetric crystal structure. *Sci. Adv.* **4**, eaar7969 (2018).

[30] M. Smidman, M. B. Salamon, H. Q. Yuan, D. F. Agterberg, Superconductivity and spin–orbit coupling in non-centrosymmetric materials: a review. *Rep. Prog. Phys.* **80**, 036501 (2017).

[31] F. Kneidinger, E. Bauer, L. Zeiringer, P. Rogl, C. Blaas-Schenner, D. Reith, R. Podloucky, Superconductivity in non-centrosymmetric materials. *Phys. C* **514**, 388-398 (2015).

[32] G. Akopov, G. Viswanathan, K. Kovnir, Synthesis, Crystal and Electronic Structure of $La_2SiP_4$, *Z. Anorg. Allg. Chem.* **647,** 91(2021).

[33] G. Akopov, J. Mark, G. Viswanathan, Shannon J. Lee, Brennan C. McBride, Juyeon Won, Frédéric A. Perras, Alexander L. Paterson, Bing Yuan, Sabyasachi Sen, Adedoyin N. Adeyemi, Feng Zhang, Cai-Zhuang Wang, Kai-Ming Ho, Gordon J. Miller, Kirill Kovnir, Third Time's the Charm: Intricate Non-centrosymmetric Polymorphism in LnSiP3 (Ln = La and Ce) Induced by Distortions of Phosphorus Square Layers, *Dalton Trans.* **50**, 6463-6476(2021).

[34] Peter Kaiser and Wolfgang Jeitschko, The Rare Earth Silicon Phosphides LnSi2P6(Ln= La, Ce, Pr, and Nd), *Journal of Solid State Chemistry* **124**, 346-352(1996).

[35] Y. S. Sun, J. D. Chen, S. D. Yang, B. X. Li, G. L. Chai, C. S. Lin, M. Luo, N. Ye, LaSiP3 and LaSi2P6: Two Excellent Rare-Earth Pnictides with Strong SHG Responses as Mid- and Far-Infrared Nonlinear Optical Crystals, *Adv. Optical Mater.* **9**, 2002176(2021).

[36] H. J. Sun, C. Zhang, W. Y. Xia, L. Tang, G. Akopov, R. H. Wang, K. M. Ho, K. Kovnir, and C. Z. Wang, Machine learning guided discovery of ternary compounds containing La, P and group IV elements, *Inorganic Chemistry* **61**, 16699-16706(2022).

[37] W. Y Xia, L. Tang, H. J. Sun, C. Zhang, K. M. Ho, G. Viswanathan, K. Kovnir, C. Z. Wang, Accelerating materials discovery using integrated deep machine learning approaches, *J. Mater. Chem. A* **11**, 25973-25982(2023).

[38] C. Chen and Shyue Ping Ong, A universal graph deep learning interatomic potential for the periodic table, *Nat. Comput. Sci.* **2**, 718–728 (2022).

[39] D. Zhang, X. Liu, X. Zhang, C. Zhang, C. Cai, H. Bi, et al., DPA-2: a large atomic model as a multi-task learner, *npj Comput. Mater.* **10**, 293 (2024).

[40] S. Batzner, A. Musaelian, L. Sun, M. Geiger, J.P. Mailoa, M. Kornbluth, N. Molinari, T.E. Smidt, B. Kozinsky, E(3)-equivariant graph neural networks for data-efficient and accurate interatomic potentials, *Nat. Commun.* **13**, 2453 (2022).

[41] H. Yang, C. Hu, Y. Zhou, X. Liu, Y. Shi, J. Li, G. Li, Z. Chen, S. Chen, C. Zeni, M. Horton, R. Pinsler, A. Fowler, D. Zügner, T. Xie, J. Smith, L. Sun, Q. Wang, L. Kong, C. Liu, H. Hao, Z. Lu, MatterSim: A Deep Learning Atomistic Model Across Elements, Temperatures and Pressures, arXiv:2405.04967 (2024).

[42] F. Shuang, Z. Wei, K. Liu, W. Gao, P. Dey, Universal machine learning interatomic potentials poised to supplant DFT in modeling general defects in metals and random alloys, arXiv:2502.03578 (2025).

[43] B.M. Wood, M. Dzamba, X. Fu, M. Gao, M. Shuaibi, L. Barroso-Luque, K. Abdelmaqsoud, V. Gharakhanyan, J.R. Kitchin, D.S. Levine, K. Michel, A. Sriram, T. Cohen, A. Das, A. Rizvi, S.J. Sahoo, Z.W. Ulissi, C.L. Zitnick, UMA: A Family of Universal Models for Atoms, Meta AI Research, https://ai.meta.com/research/publications/uma-a-family-of-universal-models-for-atoms/ (2025).

[44] Luigi Bonati and Michele Parrinello, Silicon Liquid Structure and Crystal Nucleation from Ab Initio Deep Metadynamics, *Phys. Rev. Lett.* **121**, 265701 (2018).

[45] Alessandro Laio and Michele Parrinello, *Proc. Natl. Acad. Sci. U.S.A.* **99**, 12562 (2002).

[46] A. Jain, S. P. Ong, G. Hautier, W. Chen, W. D. Richards, S. Dacek, et al., The materials project: a materials genome approach to accelerating materials innovation, *APL Mater* **1**, 011002 (2013); https://next-gen.materialsproject.org/

[47] G. Kresse, J. Furthmüller, Efficiency of ab-initio total energy calculations for metals and semiconductors using a plane-wave basis set. *Comput. Mater. Sci.* **6**, 15-50(1996).

[48] G. Kresse, J. Furthmüller, Efficient iterative schemes for ab initio total-energy calculations using a plane-wave basis set. *Phys. Rev. B* **54**, 11169-11186(1996).

[49] J. P. Perdew, K. Burke, and M. Ernzerhof, Generalized Gradient Approximation Made Simple, *Phys. Rev. Lett.* **77**, 3865(1996).

[50] S. Nosé, A unified formulation of the constant temperature molecular dynamics methods, *The Journal of Chemical Physics* **81**, 511-519(1984).

[51] W.G. Hoover, Canonical dynamics: Equilibrium phase-space distributions, *Physical Review A* **31**, 1695-1697(1985).

[52] S. Plimpton, Fast Parallel Algorithms for Short-Range Molecular Dynamics, *J. Comput. Phys.* **117**, 1(1995).

[53] J. R. Morris, C. Z. Wang, K. M. Ho, and C. T. Chan, Melting line of aluminum from simulations of coexisting phases, *Phys. Rev. B* **49**, 3109 (1994).

[54] Kathleen Lee, Saeed Kamali, Tore Ericsson, Maverick Bellard, and Kirill Kovnir, GeAs: Highly Anisotropic van der Waals Thermoelectric Material, *Chemistry of Materials* **28**, 2776-2785 (2016).

[55] ASM Alloy Phase Diagram Database, https://matdata.asminternational.org/

[56] D. Alfe and M. J. Gillan, Exchange-correlation energy and the phase diagram of Si, *Phys. Rev. B* **68**, 205212 (2003).

[57] Ling Tang, Weiyi Xia, Gayatri Viswanathan, Ernesto Soto, Kirill Kovnir, and Cai-Zhuang Wang, Synthesis challenges, thermodynamic stability, and growth kinetics of La-Si-P ternary compounds, to be published.